\documentclass[12pt]{article}

\usepackage{epsfig} 
\usepackage{amsfonts, amssymb, amsthm, graphicx}
\usepackage{amsmath, accents, bbold}
\usepackage{float}
\usepackage{authblk}
\usepackage{cmll}
\usepackage{dsfont}
\usepackage{array}
\usepackage{textgreek}
\usepackage{cite}
\usepackage{listings}
\usepackage{color}
\usepackage[colorlinks=true,allcolors=red]{hyperref}

\renewcommand{\theequation}{\thesection.\arabic{equation}}

\newcommand{\be}{\begin{equation}}
\newcommand{\ee}{\end{equation}}
\newcommand{\bea}{\begin{eqnarray}}
\newcommand{\eea}{\end{eqnarray}}
\newcommand{\dd}{\mathrm{d}}

\newcounter{orange} \renewcommand{\theorange}{\alph{orange}}

\begin{document}

\title{The continuity equation   in the phase space quantum mechanics}

\author{ Jaromir Tosiek \footnote{E-mail address: jaromir.tosiek@p.lodz.pl}}
\author{Luca Campobasso \footnote{E-mail address: luca.campobasso@p.lodz.pl }}
\affil[1]{Institute of Physics, Lodz University of Technology, ul. W\'olcza\'nska 217/221, 93-005 \L\'od\'z, Poland}
\date{\today} 
 \maketitle 

\begin{abstract}
A quantum phase space version of the continuity equation for  systems with internal degrees of freedom  is derived. 
The $1$ -- D Dirac equation is introduced and its phase space counterpart is found.
 The phase space  representation of free motion and of scattering in a nonrelativistic and  relativistic case for setups with internal degrees of freedom is discussed and illustrated. 
 Properties of Wigner functions of unbound states are analysed.

\end{abstract}

\section{Introduction}
\label{sec1}

Quantum mechanics describes the part of physical reality inaccessible to our senses. Thus, the only one way to deal with the quantum reality is mathematics. The most popular is the formalism  based on theory of  the separable Hilbert spaces. On the other hand, the classical physics is  a limit of the quantum world. Then it seems to be  natural that these two areas of science should be represented in a similar mathematical frame. Since classical mechanics in its Lagrange or Hamilton approaches works perfectly even in general relativity, we believe that the unified language of quantum and classical mechanics should be based on the phase space formalism rather than a vectorial one. Therefore we are trying to develop the so called phase space version of quantum physics. 

 The basic ingredients of this formulation of quantum theory are contained in \cite{WY31, WI32, GW46, MO49}. As milestones in development of  phase space quantum mechanics one can consider a formal series calculus \cite{bayen78}, a generalisation of the Moyal product on arbitrary symplectic manifolds \cite{fedosov94, fedosov96} and a construction of phase space for integral degrees of freedom \cite{gracia88, varilly89}. For readers interested in a holistic approach to this formalism we recommend looking into \cite{YK91, FS94, WS01, CZ05, tat, cd, lee, dit, SW07, tosiek21}. 

Our current contribution focuses on unbound states of quantum systems with internal degrees of freedom in the nonrelativistic case as well as the relativistic one. The latter option is  represented by the $1$ --D Dirac equation. Discrete degrees of freedom imply a radical change of structure of the quantum phase space. Instead of a symplectic manifold we model the system on a grid following the scheme analysed in our earlier work \cite{przanowski19}.

In some situations in order to represent scattering it is sufficient to use the continuity equation.
Thus in Sec. \ref{sec2} we derive the continuity equation for structures represented by density operators. Then we remind some facts about scattering and  introduce the phase space formalism referring to particles with discrete variables. That part is based on our earlier works \cite{przanowski14, przanowski19}. In Sec. \ref{sec5} the phase space version of nonrelativistic and relativistic continuity equation is discussed. 
 Section  \ref{sec6} contains four examples. First we analyse  a free motion of the nonrelativistic   particle and  the structure of its Wigner eigenfunction. Then we look at the $1$ -- D free motion of the Dirac particle and its nonrelativistic limit.
Presenting the phase space representation of the nonrelativistic scattering we find out more interference components of Wigner functions.
Finally  the relativistic scattering reveals the famous Klein paradox discussed for the first time in \cite{Kl29}. 

Phase space quantum mechanics is an autonomous complete formalism. Thus all of considerations contained in our article can be done exclusively in its frames. However, since calculations are sometimes very tedious,  for two systems we apply the correspondence between the Hilbert space quantum theory and its phase space counterpart. 

The  cases presented as examples are  elementary. We chose them because they illustrate very clearly opportunities offered by the phase space formulation of quantum physics.

 \section{The time evolution of spatial density of probability in the Hilbert space  quantum mechanics}
 \label{sec2}
 
The continuity equation applied to the spatial density of probability expresses the conservation of probability. It plays an important  role e.g. in analysis
of  scattering processes. The continuity equation is usually written in terms of a wave function. However, since our goal is construction of the phase space counterpart of it, we
need to  derive its more general version based on a density operator $\hat{\varrho}(t)$ first. 
  
Assume our object of interest is a quantum particle living in space ${\mathbb R}^3.$ This particle has an internal structure enabling $s+1, \, s \in {\mathcal N}$
 eigenstates. Thus the states of this particle are represented by density operators acting in the tensor product of Hilbert spaces $L^2 ({\mathbb R}^3) \otimes
{\mathbb C}^{s+1}.$

Under the aforementioned assumption the spatial density of probability $\rho({\vec r}_0,t)$ of finding the particle at a point ${\vec r}_0 \in {\mathbb R}^3$ at
instant of time $t$ in an arbitrary internal state is equal to
\[
\rho({\vec r}_0,t)={\rm Tr} \left\{ \hat{\varrho}(t)  \left(|{\vec r}_0\big> \big<{\vec r}_0|\otimes \hat{\bf 1}\right) \right\}.
 \]

The time evolution  of the density operator is  represented by the Liouville -- von Neumann equation 
\be
\label{1}
 \frac{\partial \hat{\varrho}(t)}{\partial t} + \frac{1}{i \hbar} [\hat{\varrho}(t), \hat{H}]=0.
\ee
It implies  that the speed of change of spatial density of probability satisfies the condition
\[
\frac{\partial}{ \partial t}\rho({\vec r}_0,t) =-
\frac{1}{i \hbar}  {\rm Tr} \left\{ \big[\hat{\varrho}(t), \hat{H} \big] (|{\vec r}_0\big> \big<{\vec r}_0|\otimes \hat{\bf 1})\right\}.\]
Calculating the trace in the position representation we get
\be
\label{nn28}
\frac{\partial}{ \partial t}\rho({\vec r}_0,t)=
- \frac{1}{i \hbar} \sum_{k=0}^s \int_{{\mathbb R}^3} d {\vec r} \; \big< {\vec r},k| \big[\hat{\varrho}(t), \hat{H} \big] |{\vec
r}_0\big> \big<{\vec r}_0|\otimes \hat{\bf 1}| {\vec r},k\big>
=- \frac{1}{i \hbar} \sum_{k=0}^s \big< {\vec r}_0,k|  \big[\hat{\varrho}(t), \hat{H} \big] |{\vec r}_0,k\big>.
\ee
We restrict to systems with  the Hamilton operator  represented  by sums of self -- adjoint terms $\hat{T}$ and $\hat{V}$ such that
\be
\label{dod1}
\hat{H} =\hat{T} + \hat{V}\otimes \hat{V}_{\rm int}
\ee
where
\[
[\hat{T}, \hat{\vec{p}}\,]= \hat{0}  \;\;\;
{\rm and } \;\;\;
[\hat{V}, \hat{\vec{r}}\,]= \hat{0}.
\]
 Please notice that the postulated form of Hamiltonian excludes e.g. the presence of a magnetic field or the spin -- orbit interaction. As usually, the
operator of momentum is $\hat{\vec{p}}=(\hat{p}_x,\hat{p}_y, \hat{p}_z),$ and the operator of position equals $\hat{\vec{r}}=(\hat{x},
\hat{y}, \hat{z}).$
Moreover, let $\hat{\sigma}: {\mathbb C}^{s+1} \rightarrow {\mathbb C}^{s+1} $ be an operator representing the internal degree of freedom, which eigenvectors are kets $\big| k \big>$ indexed
by $ k=0, \ldots s.$ The set of vectors $\{\big| k \big>\}_{k=0}^{s+1}$ constitutes a basis of the space $ {\mathbb C}^{s+1}.$ 
The element $\hat{V}\otimes \hat{V}_{\rm int}$ represents  potential energy. We do not specify its nature and  call it simply ``potential''. 


Since for the operator \eqref{dod1} the sum 
$
\sum_{k=0}^s \big< {\vec r}_0,k|  \big[\hat{\varrho}(t), \hat{V} \otimes \hat{V}_{\rm int} \big] |{\vec r}_0,k\big>
$
disappears, we conclude 
that the time derivative of the spatial density of probability is equal to
\be
\label{001}
\frac{\partial }{\partial t}\rho({\vec r}_0,t)= - \frac{1}{i \hbar} \sum_{k=0}^s \big< {\vec r}_0,k| [\hat{\varrho}(t), \hat{T} ]|{\vec r}_0,k\big>.
\ee
This formula resembles  very much the continuity equation
\be
\label{01}
\frac{\partial }{\partial t}\rho({\vec r}_0,t) + {\rm div} {\vec j}({\vec r}_0,t)=0.
\ee
Indeed the expression \eqref{001} would be the continuity equation
if there existed a vector ${\vec j}({\vec r}_0,t)$ such that
\be
\label{101}
{\rm div} {\vec j}({\vec r}_0,t)= \frac{1}{i \hbar} \sum_{k=0}^s \big< {\vec r}_0,k| [\hat{\varrho}(t), \hat{T} ]
|{\vec r}_0,k\big>.
\ee
For the fixed operators $\hat{\varrho}(t)$ and $\hat{T}$ one can usually introduce several vectors ${\vec j}({\vec r}_0,t)$ fulfilling that condition. In
the next two subsections we discuss the problem of choice of vector ${\vec j}({\vec r}_0,t)$ in a $3$ -- D nonrelativistic case and then for a
$1$ -- D relativistic particle satisfying the Dirac equation.

\subsection{The nonrelativistic particle in the $3$ -- D space}

Let us derive an explicit form of the continuity equation for a nonrelativistic particle when   when $\hat{T}= \frac{\hat{\vec p}^{\,2}}{2M} \otimes \hat{\bf 1}.$ 
Substituting this operator into \eqref{101} we obtain that
\be
\label{12071}
 \big< {\vec r}_0,k|  \big[ \hat{\varrho}(t), \hat{T} \big] |{\vec r}_0,k\big>
=  \frac{1}{2M} 
\big< {\vec r}_0,k| \big( \hat{\varrho}(t) \hat{\vec p} + \hat{\vec p} \hat{\varrho}(t) \big) \cdot \hat{\vec p} - \hat{\vec p} \cdot
\big(\hat{\varrho}(t) \hat{\vec p} +
\hat{\vec p}  \hat{\varrho}(t) \big) |{\vec r}_0,k\big>.
\ee
To simplify notation we have omitted the identity operator $\hat{\bf 1}.$ Applying the observations that in the position representation
$\hat{\vec p}= - i \hbar \left( \frac{\partial}{\partial x}, \frac{\partial}{\partial y}, \frac{\partial}{\partial z}\right) $ and
$\hat{\vec p}= \hat{\vec p}^{\; \dagger},$ we deduce that formula \eqref{12071} is equal to
$
- \frac{i \hbar}{2M} {\rm div} \big< {\vec r}_0,k| \hat{\varrho}(t) \hat{\vec p} + \hat{\vec p} \hat{\varrho}(t) |{\vec r}_0,k\big>.
 $
 Therefore the most natural choice of the current density vector is 
 \be
\label{011}
{\vec j}({\vec r}_0,t):= \frac{1}{2M} \sum_{k=0}^s \big< {\vec r}_0,k| \hat{\varrho}(t) \hat{\vec p} + \hat{\vec p} \hat{\varrho}(t)
|{\vec r}_0,k\big>.
\ee
 When the  system is a pure state 
 $
 \hat{\varrho}(t)= \big| \Psi(t) \big> \big< \Psi(t) \big|,
 $
 the  current density vector equals
 \be
 \label{11}
\vec{j}({\vec r}_0,t)= - \frac{\hbar}{2Mi}\sum_{k=0}^s \left(\Psi({\vec r}_0,k,t) {\rm grad} \overline{\Psi}({\vec r}_0,k,t) -
\overline{\Psi}({\vec r}_0,k,t)
 {\rm grad} \Psi({\vec r}_0,k,t)
 \right),
 \ee
 as expected. The bar $\overline{\Psi}$ stands for the complex conjugation.

 
 \subsection{The  $1$ -- D Dirac equation}
 \label{sub2.2}

In this paragraph we  consider the continuity equation for a  relativistic particle described by the $1$ -- D Dirac equation. 
Now the interpretation of
the conservation law \eqref{01} changes, because it refers to the electric charge rather than the spatial probability but the general idea of its construction
remains the same.

 At the beginning the $1$ -- D Dirac equation was treated as a toy model but now one can observe a growing interest in it (see \cite{JK20} and works quoted therein). This formula is used e.g. in modelling graphene. 

The general shape of the free Hamilton operator for the Dirac particle is
\be
\label{2.1}
\hat{H}= \alpha (c \hat{p}) + \beta  M c^2 
\ee
where $\alpha$ and $\beta$ are some square matrices. Due to the requirement that in relativistic mechanics
\[
\hat{H}^2= c^2 \hat{p}^2 + M^2 c^4 
\]
we can see that there must be
\be
\label{nn270}
\alpha^2= {\mathbf 1} \;\;, \;\; \alpha \cdot \beta + \beta \cdot \alpha = {\mathbf 0} \;\; , \;\;
\beta^2= {\mathbf 1},
\ee
where the symbols ${\mathbf 1}$ and ${\mathbf 0}$ refer to the identity matrix and the zero matrix respectively.

On the contrary to the $3$ -- D case the system of conditions \eqref{nn270} can be fulfilled by $2 \times 2$ square matrices e.g. 
 \[
 \alpha = \sigma_x = \left[\begin{array}{cc}
 0 & 1 \\
 1 & 0 
 \end{array}
 \right] \;\;\; {\rm and} \;\;\;  \beta= \sigma_z = \left[\begin{array}{cc}
 1 & 0 \\
 0 & -1 
 \end{array}
 \right]
 \]
where $\sigma_x$ and $\sigma_z$ are the respective Pauli matrices.

 The $1$ -- D  time-dependent Dirac equation is of the form
 \[
 i \hbar \frac{\partial}{\partial t}|\Psi(t) \big>= \hat{H}|\Psi(t) \big>
 \]
Vector $|\Psi(t) \big>$ is a two -- component object $|\Psi(t) \big>= \left[ \begin{array}{c}|\Psi_1(t) \big> \\
|\Psi_{0}(t) \big>\end{array}\right]$ belonging to the tensor product of Hilbert spaces $ {\mathcal H} \otimes {\mathbb C}^2, $ where
the space ${\mathcal H}$ is isomorphic to $L^2({\mathbb R}).$

 The linear space ${\mathbb C}^2$ is  spanned by eigenvectors of operator $\hat{\sigma}_z$  and its eigenvectors $|1 \rangle $,   $ |0 \rangle$ obey the equations
\be
\label{nn4}
\hat{\sigma}_z |1 \rangle = 1 \cdot |1 \rangle \;\;\; , \;\;\; \hat{\sigma}_z |0 \rangle = -1 \cdot |0 \rangle.
\ee
We apply an apparently odd notation, in which the vector referring to the eigenvalue $-1$ is denoted as  $|0 \rangle$ 
due to  compatibility with the phase space formalism introduced later.

The operators $\hat{\sigma}_x$ and $\hat{\sigma}_z$ do not commute with the Hamilton operator \eqref{2.1}
and can be alternatively written as
\[
\hat{\sigma}_x=   |0\rangle\langle1| + |1\rangle\langle 0 | \;\;\; , \;\;\; \hat{\sigma}_z= |1\rangle\langle1| - |0 \rangle\langle 0|.
\]
Eigenvalues of the Hamilton operator \eqref{2.1}  belong to the sum of separate intervals
$
E \in (-\infty, -Mc^2) \cup (Mc^2, + \infty).
$

Since the operators $\hat{H}$ and $\hat{p}$ commute, the eigenvalues of energy $E$ can be expressed as functions of eigenvalues of the momentum  ${\tt p} \in {\mathbb R}.$ Indeed, 
  the dispersion relation is satisfied
\begin{equation}
    E^2_{\pm} = (c{\tt p})^2 + (Mc^2)^2,\; {\rm where} \;\; E_\pm = \pm\sqrt{(c {\tt p})^2 + (Mc^2)^2}.
\end{equation}

The normalised eigenfunctions of the relativistic free particle are indexed by the value of momentum and the sign of energy  
\begin{equation}
    \psi_{{\tt p} \pm} (x) =  \frac{1}{\sqrt{2\pi \hbar}}\begin{pmatrix}
    \frac{c{\tt p}}{E_{\pm}- Mc^2} \\ 1
    \end{pmatrix}\left(\frac{(c{\tt p})^2}{(E_{\pm}- Mc^2)^2} + 1\right)^{-\frac{1}{2}} \exp \left(\frac{i{\tt p}x}{  \hbar} \right).
\end{equation}
They fulfil the  standard orthonormality conditions 
\[
   \int_\mathbb{R} \psi^{\dagger}_{{\tt p'}+}(x)\psi_{{\tt p}+}(x)\dd x =  \delta({\tt p}-{\tt p'})
   \;\; , \;\;
   \int_\mathbb{R} \psi^{\dagger}_{{\tt p'}-}(x)\psi_{{\tt p}-}(x)\dd x =  \delta({\tt p}- {\tt p'})   
\]
together with
\be
\int_\mathbb{R} \psi^{\dagger}_{{\tt p'}+}(x)\psi_{{\tt p}-}(x)\dd x =0.
\ee
We observe that in the $1$ -- D case of the Dirac equation there is no spin and solutions are parametrised exclusively by the value of momentum ${\tt p}$
and the sign of energy. {\it Indeed, there is no nontrivial operator represented by a $2 \times 2$ matrix, which would commute with the
momentum and the Hamilton operators}. A physical explanation of the phenomenon is that the spin is related to a rotation which in the $1$
-- D case has no sense.

Let us consider the $1$ -- D Dirac equation with a potential
\be
\label{nn5}
\hat{H}=  c \hat{p}  \sigma_x +  Mc^2 \sigma_z + \hat{V}(x) \otimes \hat{V}_{\rm int}.
\ee
The density operator for a $1$ -- D Dirac particle is represented as  a $2 \times 2$ matrix
 \[
 \hat{\varrho}(t)= \left[\begin{array}{cc}
 \hat{\varrho}_{11}(t) & \hat{\varrho}_{1 0}(t) \\
 \hat{\varrho}_{01}(t) & \hat{\varrho}_{00}(t)
 \end{array}
 \right] .
 \]
 The time evolution of the density operator in the Schroedinger picture is given by the general formula \eqref{1}.
 The projection operator on the state representing a particle localised in the configuration space at a point $x$  equals
 \[
 \hat{\Pi}_{x}=
| x \rangle \langle x | \otimes \hat{\mathbf 1}
=| x, 1 \rangle \langle x,1 | + | x, 0 \rangle \langle x,0 |= 
 \left[ 
 \begin{array}{cc}
  | x \rangle \langle x | & 0 \\
 0 &  | x \rangle \langle x | 
 \end{array}
 \right].
 \]
 Hence the electric charge density  at a point $x_0$ is calculated as
 \[
 \rho(x_0,t) = {\rm Tr } \left\{ q\hat{\varrho}(t) \hat{\Pi}_{x_0} \right\} 
 =  q \langle x_0 | \hat{\varrho}_{11}(t) | x_0 \rangle + q\langle x_0 | \hat{\varrho}_{00}(t) | x_0 \rangle,
 \]
where by ``$q$'' we denote the charge of the Dirac particle.
The time evolution of the charge density $\rho(x,t)$ is given by the expression
\be
\label{201}
\frac{\partial \rho(x_0,t)}{\partial t}= q \big< x_0 |\frac{\partial \hat{\varrho}_{11}(t)}{\partial t} | x_0 \big> + q \big< x_0
|\frac{\partial \hat{\varrho}_{00}(t)}{\partial t} | x_0 \big>.
\ee
 Neither the potential nor  the term $Mc^2 \sigma_z$  influence the time evolution of the spatial density of probability. Therefore  calculating 
the commutator 
$
[\hat{\varrho}(t), c\hat{p}\sigma_x]
$
from \eqref{1}  one  can see easily  that
\[
 \frac{\partial  \hat{\varrho}_{11}(t)}{\partial t}
=- \frac{c}{i \hbar} \Big(  \hat{\varrho}_{10}(t)\hat{p}- \hat{p} \hat{\varrho}_{01}(t)\Big)
\;\;\;
{\rm and}\;\;\;
\frac{\partial  \hat{\varrho}_{00}(t)}{\partial t}
 = -  \frac{c}{i \hbar} \Big(  \hat{\varrho}_{01}(t)\hat{p}- \hat{p} \hat{\varrho}_{10}(t)\Big).
\]
Relation \eqref{201} may be transformed into  a   continuity equation.
Indeed one has to   postulate that
\[
{\rm div }\vec{j}(x_0,t)= \frac{qc}{i \hbar} \langle x_0 | - \hat{p} (\hat{\varrho}_{10}(t)+\hat{\varrho}_{01}(t) ) +
(\hat{\varrho}_{10}(t)+\hat{\varrho}_{01}(t) ) \hat{p} | x_0 \rangle
\]
so the charge current density at $x_0$ equals
\be
\label{130722}
\vec{j}(x_0,t)= qc \langle x_0 |  \hat{\varrho}_{10}(t)+\hat{\varrho}_{01}(t)  | x_0 \rangle.
\ee

For a pure state $ \left[ \begin{array}{c}|\Psi_1(t) \big> \\
|\Psi_{0}(t) \big>\end{array}\right]$
 the unique component of current density  is expressed as
\be
\label{190923}
j(x_0,t)=qc(  \Psi_1(x_0,t)\overline{\Psi_{0}(x_0,t)} + \Psi_{0}(x_0,t)\overline{\Psi_1(x_0,t)}).
\ee
In that situation  the spatial charge density at the point $x_0$ is given by the relation
\[
{\rho}(x_0,t)= q \Big( |\Psi_1(x_0,t)|^2 + |\Psi_{0}(x_0,t)|^2 \Big)
\]
as expected.

\vspace{0.5cm}
At the end of this section we present a few general remarks about the current density for stationary states.  In this case at every point ${\vec r}_0$ the time derivative of the spatial density of probability  charge density  vanishes
 $
  \frac{\partial }{\partial t}\rho({\vec r}_0,t)=0
 $  
 and the continuity equation implies that
$
 {\rm div} {\vec j}(\vec{r}_0,t)=0.
$
Applying the nonrelativistic definition \eqref{011} as well as the relativistic one \eqref{130722} we observe that the partial derivative $ \frac{\partial
}{\partial t} {\vec j}(\vec{r}_0,t)=0$ so in every stationary case the current density is independent from the time coordinate. 

Moreover, since for every instant of time the equality $
 {\rm div} {\vec j}(\vec{r}_0)=0
$ is satisfied, from the Gauss theorem we deduce that even at points of discontinuity of the potential the current density is continuous.


 \section{Scattering on a potential barrier}
 
 \setcounter{equation}{0}

Theory of scattering is the part of quantum mechanics in which the continuity equation is widely used. 
In this section we present  general remarks about the systems in which scattering is observed. We focus on  cases   with   the potentials of the type $\hat{V}\otimes \hat{V}_{\rm int}$  constant at infinity i.e. such that in the position representation 
the limit
\be
\label{02}
\lim_{|\vec{r}\,|\rightarrow \infty}{\rm grad} V(\vec{r}\,)=0.
\ee
We assume that neither of the operators $\hat{V}$ and  $ \hat{V}_{\rm int}$ depends on time.
Thus
the process of scattering of particles on  the potential barrier  is stationary. 

The  complete characterisation of  scattering is done by the $S$ matrix.  However, in cases discussed in our paper it is sufficient to   use  the coefficient 
of transmission $T$ and the coefficient of reflection $R.$  

 In the  situation, when the potential $V(\vec{r}\,)$ depends exclusively on one Cartesian coordinate $x$ varying from $- \infty$ to $+ \infty$ and the source is homogeneous, these coefficients take the form
 \[
 T= \frac{|\vec{j}_{trans}(x_2)|}{|\vec{j}_{inc}(x_1)|} \;\; , \;\;R = \frac{|\vec{j}_{ref}(x_1)|}{|\vec{j}_{inc}(x_1)|}. 
 \] 
where the applied symbols mean respectively: $\vec{j}_{trans}(x_2)$ -- the  current density of transmitted particles, $\vec{j}_{ref}(x_1)$ -- the 
current density of reflected flow and finally $\vec{j}_{inc}(x_1)$ -- the  current density of incoming beam. The beams $\vec{j}_{inc}(x_1)$ and $\vec{j}_{ref}(x_1)$ are measured at an arbitrary  point $x_1$ in front of  the barrier and $\vec{j}_{trans}(x_2)$ at some point $x_2$ behind the barrier. From the continuity equation we deduce that
 \be
\label{301}
 |\vec{j}_{trans}(x_2)|=|\vec{j}_{inc}(x_1)| - |\vec{j}_{ref}(x_1)|
 \ee
 so the sum of coefficients yields the equality $T+R=1.$
 
 \begin{figure}
 \includegraphics{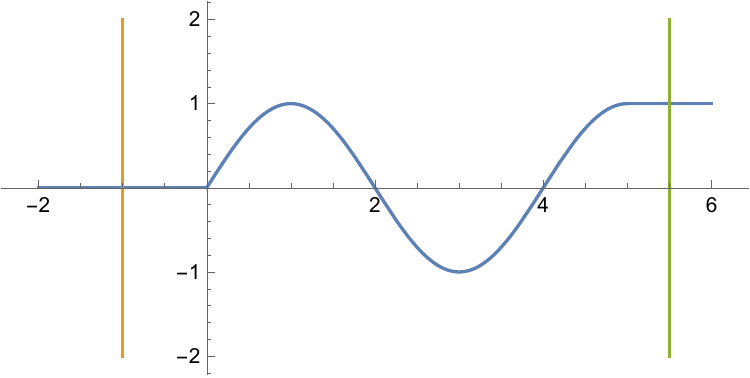}
 \caption{An example of a $1$ -- D potential fulfilling condition \eqref{02}}
 \label{wyk1}
 \end{figure}
 
Let us look at Fig. \ref{wyk1}. The sketched potential  satisfies the  conditions 
\[
{\rm for}\; x<x_L\; V(x)=0 \;\;\; {\rm and} \;\;\; {\rm for}\; x>x_R\;  V(x)=V_0
\]
where $x_L, x_R$ are the boundaries of the potential barrier. This potential can be treated as one dimensional or living in a more dimensional space but depending exclusively on one Cartesian variable $x.$

We assume that the source of particles is localised at minus infinity and their energy exceeds the value of potential $V_0.$
In both the nonrelativistic and the relativistic case
 at the left -- hand side of the barrier ($x<x_L$) a general solution of the stationary Schroedinger or the Dirac equation is a linear combination of functions
\setcounter{orange}{1}
\renewcommand{\theequation} {\arabic{section}.\arabic{equation}\theorange}
 \be
\label{a}
\Psi(x)= \left[
\begin{array}{c}A_s \\ A_{s-1} \\ \vdots \\ A_0
\end{array}
\right]
 \exp\left( \frac{i{\tt p}x}{\hbar} \right) + 
 \left[
\begin{array}{c}B_s \\ B_{s-1} \\ \vdots \\ B_0
\end{array}
\right]
  \exp\left(- \frac{i{\tt p}x}{\hbar} \right)
 \ee
referring to the eigenvalue of energy $E=\frac{{\tt p}^2}{2M}$ or $E =  \sqrt{{\tt p}^2c^2 + M^2 c^4}$ respectively 
 and at the right -- hand side of the barrier  $(x>x_R)$ we obtain
\addtocounter{orange}{1}
\addtocounter{equation}{-1}
 \be
\label{b}
 \Psi(x)=\left[
\begin{array}{c}C_s \\ C_{s-1} \\ \vdots \\ C_0
\end{array}
\right] \exp\left(\frac{i\tilde{\tt p}x}{\hbar} \right), 
 \ee
where $E=\frac{\tilde{\tt p}^2}{2M}+V_0$ or $E =  \sqrt{\tilde{\tt p}^2 c^2 + M^2 c^4}+ V_0$ respectively. 

In the nonrelativistic case the parametre $s$ usually refers to the number of possible projections of spin of the particle. For the $1$ -- D Dirac equation $s=1$ and  is related to the discrete degree of freedom particle -- antiparticle.

\renewcommand{\theequation} {\arabic{section}.\arabic{equation}}

At the left -- hand side of the barrier the total current density decomposes
into the sum of two current densities $\vec{j}_{inc}(x_1)$ and $ \vec{j}_{ref}(x_1)$ determined by functions $ \exp\left( \frac{i{\tt p}x}{\hbar}
\right)$ and $ \exp\left(- \frac{i{\tt p}x}{\hbar} \right)$ respectively. The transmitted current density $\vec{j}_{trans}(x_2)$ in the nonrelativistic case as well as in the relativistic one follows  exclusively from function
\eqref{b}.

The pure eigenstate of energy $E$ in a basis free form can be decomposed into the sum of four elements  
\be
\label{o1}
|\Psi \big> = |\Psi_{inc} \big> + |\Psi_{ref} \big>+|\Psi_{barrier} \big> + |\Psi_{trans} \big>
\ee
As the indices suggest, the ket $|\Psi_{inc} \big>$ refers to the incoming beam, $|\Psi_{ref} \big>$ contains information about  reflected particles and $|\Psi_{trans} \big>$ is the transmitted component of the state $|\Psi \big>.$
The vector $|\Psi_{barrier} \big>$ represents the particles interacting with the barrier and it does not come directly into the coefficients $R$ and $T.$

These vectors satisfy the following orthogonality conditions:
\be
\label{o1.1}
\left(\big<\Psi_{inc} | + \big<\Psi_{ref} | \right)|\Psi_{barrier} \big>=0\; , \; \left(\big<\Psi_{inc} | + \big<\Psi_{ref} | \right)|\Psi_{trans} \big>=0 \; , \; \big<\Psi_{barrier} | \Psi_{trans} \big>=0.
\ee
The density operator equals of the pure state $| \Psi \rangle$ equals
\be
\label{o0}
\hat{\varrho}
=\Big(|\Psi_{inc} \big> + |\Psi_{ref} \big>+|\Psi_{barrier} \big> + |\Psi_{trans} \big>\Big)
\Big(\langle \Psi_{inc} | + \langle \Psi_{ref} |+ \langle \Psi_{barrier} | + \langle \Psi_{trans} |\Big).
\ee

Let us see which components of the density operator of the state \eqref{o1}
really
contribute to the current density. We start from the nonrelativistic case so the current density of probability is given by formula \eqref{011}.

At the point  $x_1$ at the LHS of the barrier we can see that due to the spatial separation between functions $\langle x|\Psi_{inc}\rangle + \langle
x|\Psi_{ref}\rangle$, $ \langle x|\Psi_{barrier}\rangle$ and $ \langle x|\Psi_{trans} \rangle$ in formula \eqref{011} only four elements
survive. Thus the density current of probability at $x_1$ is calculated as
\[
\vec{j}(x_1)= \frac{1}{2M} \sum_{k=0}^{s}\Big(  \langle x_1,k| \Psi_{inc} \rangle \langle \Psi_{inc} | \hat{p}| x_1,k \rangle 
+\langle x_1,k|\hat{p} |\Psi_{inc} \big> \langle \Psi_{inc} |  x_1,k \rangle 
\]
\[
+
\langle x_1,k| \Psi_{ref} \big> \langle \Psi_{ref} | \hat{p}| x_1,k \rangle 
+\langle x_1,k|\hat{p} |\Psi_{ref} \big> \langle \Psi_{ref} |  x_1,k \rangle
\]
\[
+
\langle x_1,k| \Psi_{inc} \big> \langle \Psi_{ref} | \hat{p}| x_1,k \rangle 
+\langle x_1,k|\hat{p} |\Psi_{inc} \big> \langle \Psi_{ref} |  x_1,k \rangle 
\]
\be
\label{o2}
+
\langle x_1,k| \Psi_{ref} \big> \langle \Psi_{inc} | \hat{p}| x_1,k \rangle 
+\langle x_1,k|\hat{p} |\Psi_{ref} \big> \langle \Psi_{inc} |  x_1,k \rangle \Big).
\ee
Moreover, the components  $| \Psi_{ref} \big> \langle \Psi_{inc} |$ and $| \Psi_{inc} \big> \langle \Psi_{ref} |$ do
not give a contribution to the current density $\vec{j}(x_1).$ 
Finally
 the incoming current density equals
\[
\vec{j}_{inc}(x_1):= \frac{1}{2M} \sum_{k=0}^{s}\Big(  \langle x_1,k| \Psi_{inc} \rangle \langle \Psi_{inc} | \hat{p}| x_1,k \rangle 
+\langle x_1,k|\hat{p} |\Psi_{inc} \big> \langle \Psi_{inc} |  x_1,k \rangle \Big)
\]
and the reflected current density is given by formula
\[
\vec{j}_{ref}(x_1):= \frac{1}{2M} \sum_{k=0}^{s}\Big( \langle x_1,k| \Psi_{ref} \big> \langle \Psi_{ref} | \hat{p}| x_1,k \rangle 
+\langle x_1,k|\hat{p} |\Psi_{ref} \big> \langle \Psi_{ref} |  x_1,k \rangle \Big).
\]
At the point $x_2$ at  the right-hand side  the barrier depicted at Fig. \ref{wyk1} we obtain that the current density of probability is equal to 
\be
\label{o4}
\vec{j}(x_2)=\vec{j}_{trans}(x_2):= \frac{1}{2M} \sum_{k=0}^{s}\Big( \langle x_2,k| \Psi_{trans} \big> \langle \Psi_{trans} | \hat{p}|
x_2,k \rangle
+\langle x_2,k|\hat{p} |\Psi_{trans} \big> \langle \Psi_{trans} |  x_2,k \rangle \Big).
\ee
We get $\vec{j}_{trans}(x_2)$ directly by application of expression \eqref{011} to the density operator \eqref{o0}. 
Summing up  we can see that the current densities $\vec{j}_{trans}(x), \; \vec{j}_{ref}(x)$ and $\vec{j}_{inc}(x)$
are completely determined by the following components of the density
operator: $|\Psi_{inc} \big> \big<\Psi_{inc} |, $
$|\Psi_{ref} \big> \big<\Psi_{ref} | $ and $|\Psi_{trans} \big> \big<\Psi_{trans} |. $ 

The analogous observation is valid for the $1$ -- D Dirac equation.

Technically the way to distinguish between the projection operator  $|\Psi_{inc} \big> \big<\Psi_{inc} | $ and $|\Psi_{ref} \big> \big<\Psi_{ref} | $ is not trivial.  The reason is that the incoming wave and
the reflected one are not mutually orthogonal as
\[
\langle \Psi_{ref}|\Psi_{inc}\rangle \sim 
  - i \exp \left( \frac{2ix_{L}{\tt p}}{\hbar}\right) \,{\rm vp} \frac{1}{\tt  p} + \pi\,\delta({\tt p}). 
 \]
 
 Let us introduce a projection operator
\be
\label{1101}
\hat{I}_{-}:= \sum_{k=0}^s\int_{- \infty}^{x_L} |x,k \rangle dx \langle x,k|.
\ee
Then for the density operator \eqref{o0} the equality holds
\be
\label{1102}
\hat{I}_{-} \hat{\varrho} \hat{I}_{-}= |\Psi_{ref} \big> \langle \Psi_{ref} | + |\Psi_{inc} \big> \langle \Psi_{inc} |+ |\Psi_{inc}
\big> \langle \Psi_{ref} |+ |\Psi_{ref} \big> \langle \Psi_{inc} | .
\ee
We discuss the nonrelativistic case first.

Since the density operator \eqref{o0} represents projection on the eigenstate of the Hamilton operator $\hat{H},$ referring to the
eigenvalue $\sqrt{2ME},$  the  limit
\be
\label{1103a}
\lim_{G \rightarrow + \infty}\frac{1}{G} \sum_{k=0}^s \int_{- G}^{+ \infty } \langle x,k | \hat{I}_{-} \sqrt{2M\hat{H}} \hat{\varrho}
\hat{I}_{-} | x, k \rangle dx= 2 {\tt p} \sum_{k=0}^s \Big( |A_k|^2 + |B_k|^2 \Big).
\ee
On the other hand  we know that 
\be
\label{1.77}
|\vec{j}_{inc}(x_1)|= \frac{2{\tt p}}{M}  \sum_{k=0}^s |A_k|^2 \;\;\; {\rm and}\;\;\; |\vec{j}_{ref}(x_1)|=  \frac{2 {\tt p}}{M}  \sum_{k=0}^s |B_k|^2. 
\ee

Therefore  the incoming current density of probability equals
\be
\label{1104}
\vec{j}_{inc}(x_1)= \frac{1}{2M}\lim_{G \rightarrow + \infty}\frac{1}{G} \sum_{k=0}^s \int_{- G}^{+ \infty } \langle x,k | \hat{I}_{-}
\sqrt{2M\hat{H}}\hat{\varrho} \hat{I}_{-} | x, k \rangle dx +\frac{\vec{j}(x_1)}{2}
\ee
and the reflected  current density of probability can be calculated as
\be
\label{1105}
\vec{j}_{ref}(x_1)= -\frac{1}{2M}\lim_{G \rightarrow + \infty}\frac{1}{G} \sum_{k=0}^s \int_{- G}^{+ \infty } \langle x,k | \hat{I}_{-}
\sqrt{2M\hat{H}} \hat{\varrho} \hat{I}_{-} | x, k \rangle dx + \frac{\vec{j}(x_1)}{2}.
\ee

 If one  proposes a self -- adjoint operator
\[
\widehat{|p|}_{-}:=\frac{1}{2} \left( \hat{I}_{-}\sqrt{2M\hat{H}} + \sqrt{2M\hat{H}} \hat{I}_{-} \right),
\]
then the term \eqref{1103a} represents the mean value of $\widehat{|p|}_{-}$. Thus alternatively
\be
\label{1104a}
\vec{j}_{inc}(x_1)= \frac{\langle \widehat{|p|}_{-} \rangle }{2M} +\frac{\vec{j}(x_1)}{2}
\;{\rm
and}\;\;
\vec{j}_{ref}(x_1)= - \frac{\langle \widehat{|p|}_{-} \rangle }{2M} +\frac{\vec{j}(x_1)}{2}.
\ee
Notice that the quantities $\frac{\langle \widehat{|p|}_{-} \rangle }{2M}$ and $\frac{\vec{j}(x_1)}{2}$ do not depend on a choice of the point  $x_1 < x_L$.

The case of $1$ --D Dirac equation is similar. Instead of operator $ \hat{I}_{-}
\sqrt{2M\hat{H}}\hat{\varrho} \hat{I}_{-}$ in formula \eqref{1103a} we simply put  
\be
\label{1104b}
\frac{1}{c}\hat{I}_{-}
\sqrt{\hat{H}^2-M^2c^4}\,\hat{\varrho} \hat{I}_{-}
\ee
and then follow the path proposed above.

 
 \section{The phase space formulation of quantum theory}

\setcounter{equation}{0}

In this section we remind some basic facts about  phase space quantum mechanics.  This attempt to quantum theory  is
equivalent to its Hilbert space version and can be applied without any references to it. However, since  quantum internal
degrees of freedom do not have their classical counterparts, it is much easier to use a bridge between phase space quantum mechanics
and the Hilbert space formulation in order to construct their phase space representation. Ideas described below were presented in full in
\cite{ przanowski19, tosiek21, przanowski17}.

The quantum system under consideration is modelled on the Hilbert space
$
L^2({\mathbb R}^3) \otimes {\mathbb C}^{s+1}.
$
In the set of square integrable functions $L^2({\mathbb R}^3)$  the canonically conjugated  operators of position
$\hat{\vec{r}}$ and  of momentum $\hat{\vec{p}}$ act.
Using these basic operators we build a family of  unitary operators called displacement operators  
\[
 \widehat{\mathcal{U}}(\vec{\lambda},\vec{\mu}):=\exp\{i(\vec{\lambda}\cdot\hat{\vec{p}}+\vec{\mu}\cdot\hat{\vec{r}}\,)\} 
\]
\be
\label{402}
= \exp\left\{-i \frac{\hbar \vec{\lambda}\cdot
\vec{\mu}}{2}\right\}\exp\{i\vec{\lambda}\cdot\hat{\vec{p}}\,\}\exp\{i\vec{\mu}\cdot\hat{\vec{r}}\,\}
= \exp\left\{i \frac{\hbar \vec{\lambda}\cdot \vec{\mu}}{2}\right\}\exp\{i \vec{\mu}\cdot\hat{\vec{r}}\,\}\exp\{i
\vec{\lambda}\cdot\hat{\vec{p}}\,\}
\ee
where vectors
$\vec{\lambda}, \; \vec{\mu} \in {\mathbb R}^3$,
 and the symbol ``$\cdot$'' denotes the scalar product.

Alternatively the displacement operators can be written as
\be
\label{215}
\widehat{\mathcal{U}}(\vec{\lambda},\vec{\mu})= \int\limits_{\mathbb{R}^{3}}\exp\{i\vec{\mu}\cdot
\vec{r}\,\}\Big|\vec{r}-\frac{\hbar\vec{\lambda}}{2}\Big> d\vec{r} \,\Big< \vec{r}+\frac{\hbar\vec{\lambda}}{2}\Big|
= \int\limits_{\mathbb{R}^{3}}\exp\{i\vec{\lambda}\cdot \vec{p}\,\}\Big| \vec{p}+\frac{\hbar\vec{\mu}}{2}\Big> d\vec{p} \,\Big<
\vec{p}-\frac{\hbar \vec{\mu}}{2} \Big|.
\ee

In the finite dimensional Hilbert space ${\mathbb C}^{s+1}$ we do not have any pair of canonically conjugated operators.  Thus we start from an arbitrary orthonormal basis $\{|n\rangle\}^{s}_{n=0}$ and build another complete orthonormal
system of vectors $\{|\phi_{m}\rangle\}^{s}_{m=0}$ according to the rule
\[
 |\phi_{m}\rangle\:=\frac{1}{\sqrt{s+1}}\sum_{n=0}^{s}\exp\{i n \phi_{m}\} |n\rangle, \,\,\, m=0,...,s
\]
where the numbers 
$
 \phi_{m}=\frac{2\pi}{s+1}m, \,\,\, m=0,...,s.
$

With the use of  sets of   projection operators 
$ \{
|n\rangle\langle n| \}_{n=0}^s$  and $ \{|\phi_{m}\rangle\langle \phi_{m}|\}_{ m=0}^s
$
we introduce two self -- adjoint operators
\begin{equation}\label{25}
  \hat{n}:=\sum_{n=0}^{s}n|n\rangle\langle n|\,\,\, , \,\,\, \hat{\phi}:=\sum_{m=0}^{s}\phi_{m}|\phi_{m}\rangle\langle \phi_{m}|,
\end{equation} 
 then the so called Schwinger operators 
\begin{equation}\label{26}
  \hat{V}:=\exp\left\{i \frac{2\pi}{s+1}\hat{n}\right\}, \,\,\, \hat{R}:=\exp\{i\hat{\phi}\}
\end{equation}
and finally a
 family of unitary operators 
\begin{equation}\label{29}
  \hat{\mathcal{D}}(k,l):=\exp\left\{-i\frac{\pi k l}{s+1}\right\}\hat{R}^{k}\hat{V}^{l}, \,\,\, k,l \in \mathbb{Z}
\end{equation}
parametrised by the integers $k$ and $l$ and playing in the Hilbert space ${\mathbb C}^{s+1}$ the role analogous to the displacement operators \eqref{402}
in the space $L^2({\mathbb R}^3).$

The displacement operators $\hat{\mathcal{U}}(\vec{\lambda},\vec{\mu})$ together with $\hat{\mathcal{D}}(k,l)$ enable us to establish a one -- to -- one
relationship between the Hilbert space version and a phase space formulation of quantum mechanics.
 The counterpart of the Hilbert space $ L^2({\mathbb R}^3) $ is well
known. This is the classical symplectic space ${\mathbb R}^3 \times {\mathbb R}^3$ of the system. The internal states are represented on
the $(s+1) \times (s+1)$ grid being a discrete phase space denoted as $\Gamma^{s+1}$. Thus the phase space quantum mechanics is built on the
set
\[
\Gamma={\mathbb R}^3 \times {\mathbb R}^3 \times \Gamma^{s+1} 
\]
 and the coordinates of points belonging to $\Gamma$ are $(\vec{p},\vec{r},\phi_m,n).$

A correspondence between a linear operator $\hat{f}$ acting in the Hilbert space $
L^2({\mathbb R}^3) \otimes {\mathbb C}^{s+1}$ and its respective function $f(\vec{p},\vec{r},\phi_{m},n)$ on the space $\Gamma$ is
established as
\[
f(\vec{p},\vec{r},\phi_{m},n)=\left(\frac{\hbar}{2\pi}\right)^{3}(s+1)^{-1}\sum_{k,l=0}^{s}\int_{\mathbb{R}^{3}\times\mathbb{R}^{3}}d\vec{\lambda}
d\vec{\mu}\left(\mathcal{P}\left(\frac{\hbar \lambda_x\mu_x}{2}, \frac{\hbar \lambda_y\mu_y}{2}, \frac{\hbar
\lambda_z\mu_z}{2}\right)\mathcal{K}\left(\frac{\pi k l}{s+1}\right)\right)^{-1}
\]
\be
\label{231}
\exp\{i(\vec{\lambda}\cdot \vec{p} +\vec{\mu}\cdot \vec{r}\,)\}\exp\left\{ i\frac{2\pi}{s+1}(km+ln)\right\}
\textrm{Tr}\left\{\hat{f}\hat{\mathcal{U}}^{\dag}(\vec{\lambda},\vec{\mu})\hat{\mathcal{D}}^{\dag}(k,l)\right\} 
\ee
and it depends on two additional functions $\mathcal{P} \left(\frac{\hbar \lambda_x\mu_x}{2}, \frac{\hbar \lambda_y\mu_y}{2}, \frac{\hbar
\lambda_z\mu_z}{2}\right)$ and $\mathcal{K}\left(\frac{\pi k l}{s+1}\right)$ known as kernels.
On the  component ${\mathbb R}^3 \times {\mathbb R}^3$ of the phase space we  choose the Weyl ordering so  we put
$
\mathcal{P} \left(\frac{\hbar \lambda_x\mu_x}{2}, \frac{\hbar \lambda_y\mu_y}{2}, \frac{\hbar \lambda_z\mu_z}{2}\right)=1.
$
A decision about selecting the $\mathcal{K}\left(\frac{\pi k l}{s+1}\right)$ is not so simple because one cannot postulate that
$\mathcal{K}\left(\frac{\pi k l}{s+1}\right)=1$ (see \cite{przanowski17}). Since in our further considerations the parameter $s=1,$ the best admissible option seems to be
$
\mathcal{K}\left(\frac{\pi k l}{s+1}\right)=(-1)^{kl}.
$

Defining the family of operators called the Fano operators or the Stratonovich -- Weyl quantiser
$$
\widehat{\Omega}[\mathcal{P},\mathcal{K}](\vec{p},\vec{r},\phi_{m},n):=\left(\frac{\hbar}{2\pi}\right)^{3}(s+1)^{-1}\sum_{k,l=0}^{s}\int_{\mathbb{R}^{3}\times\mathbb{R}^{3}}d\vec{\lambda}
\,d\vec{\mu}\,
\mathcal{P} \left(\frac{\hbar \lambda_x\mu_x}{2}, \frac{\hbar \lambda_y\mu_y}{2}, \frac{\hbar
\lambda_z\mu_z}{2}\right)\mathcal{K}\left(\frac{\pi k l}{s+1}\right)
$$
\begin{equation}\label{232}
\times \exp\{-i(\vec{\lambda}\cdot \vec{p} +\vec{\mu}\cdot \vec{r})\}\exp\left\{-i\frac{2\pi}{s+1}(km+ln)\right\}
\hat{\mathcal{U}}(\vec{\lambda},\vec{\mu})\hat{\mathcal{D}}(k,l)
\end{equation}
we  reduce the correspondence rule \eqref{231} to a compact form
\be
\label{2321}
f(\vec{p},\vec{r},\phi_{m},n)= \textrm{Tr}\left\{ \hat{f} \, \hat{\Omega}[\mathcal{P},\mathcal{K}](\vec{p},\vec{r},\phi_{m},n) \right\}.\ee

A phase space counterpart of the density operator $\hat{\varrho}$ is known as the Wigner function $W(\vec{p},\vec{r},\phi_{m},n).$ For
the choice of kernels proposed above we introduce the Wigner function via formula
\be
\label{2322}
W(\vec{p},\vec{r},\phi_{m},n,t)= \frac{1}{(2\pi\hbar)^{3}(s+1)} \textrm{Tr}\left\{\hat{\varrho}(t)\,
\hat{\Omega}[1,(-1)^{kl}](\vec{p},\vec{r},\phi_{m},n)\right\}.
\ee
Thus the mean value of an observable represented by a function $f(\vec{p},\vec{r},\phi_{m},n)$ equals
\begin{equation}
\label{44}
\langle f(\vec{p},\vec{r},\phi_{m},n) \rangle(t)=\sum_{m,n=0}^{s}\int_{\mathbb{R}^{3}\times\mathbb{R}^{3}}d\vec{p} \,d\vec{r}\,
f(\vec{p},\vec{r},\phi_{m},n) W(\vec{p},\vec{r},\phi_{m},n,t).
\end{equation}

 The time evolution of Wigner function is given by the Liouville -- von Neumann -- Wigner equation
 \be
\label{x2}
\frac{\partial}{\partial t}W(\vec{p},\vec{r},\phi_{m},n,t) + \{W(\vec{p},\vec{r},\phi_{m},n,t), H(\vec{p},\vec{r},\phi_{m},n,t)\}_M=0, \ee
  where the Moyal bracket is calculated as
 \[
\{W(\vec{p},\vec{r},\phi_{m},n,t), H(\vec{p},\vec{r},\phi_{m},n,t)\}_M
\]
\[
:= \frac{1}{i \hbar} \Big(W(\vec{p},\vec{r},\phi_{m},n,t) \ast H(\vec{p},\vec{r},\phi_{m},n,t)- H(\vec{p},\vec{r},\phi_{m},n,t) \ast
W(\vec{p},\vec{r},\phi_{m},n,t) \Big).
 \]
By $H(\vec{p},\vec{r},\phi_{m},n,t)$ the Hamilton function is denoted. The asterix ``$\ast$'' symbolises the famous star product. 
For the Weyl ordering and the discrete kernel $\mathcal{K}\left(\frac{\pi k l}{s+1}\right)=(-1)^{kl}$ when one assumes that $s=1$
the star product of functions is equal to
\begin{multline}\label{610}
(f*g)(\vec{p},\vec{r},\phi_{m},n,t)=\frac{4}{(2\pi\hbar)^{6}}\sum_{m',n', m'',n''=0}^{1}
\int_{\mathbb{R}^{12}} d\vec{p}\,' d\vec{r}\,' d\vec{p}\,'' d\vec{r}\,'' \\
\times  f(\vec{p}\,',\vec{r}\,',\phi_{m'},n',t) g(\vec{p}\,'',\vec{r}\,'',\phi_{m''},n'',t)
\exp\left\{\frac{2i}{\hbar}[(\vec{r}-\vec{r}\,')\cdot(\vec{p}-\vec{p}\,'')-(\vec{r}-\vec{r}\,'')\cdot(\vec{p}-\vec{p}\,')]\right\} \\
\times
\left\{(1+(-1)^{m'+m''})(1+(-1)^{n'+n''})+(-1)^{m}((-1)^{m'}+(-1)^{m''})+(-1)^{m+n}((-1)^{m'+n'}+(-1)^{m''+n''})\right.\\
+(-1)^{n}((-1)^{n'}+(-1)^{n''})+i\left[(-1)^{m}(-1)^{n'+n''}((-1)^{m'}-(-1)^{m''})\right.\\
\left.\left.+(-1)^{m+n}((-1)^{m''+n'}-(-1)^{m'+n''})+(-1)^{n}(-1)^{m'+m''}((-1)^{n''}-(-1)^{n'})\right]\right\}.
\end{multline}


\section{The continuity equation in the phase space quantum mechanics}
\label{sec5}

\setcounter{equation}{0}

In order to find a phase space counterpart of the continuity equation we use the following marginal distribution \cite{przanowski19}
 \[
\sum_{m,n=0}^{s} \int_{{\mathbb R}^3} d\vec{p}\, W(\vec{p},\vec{r},\phi_{m},n,t)= {\rm Tr} \left( \hat{\varrho}(t) |{\vec r}\big>
\big<{\vec r}|\otimes \hat{\mathbf 1}\right)=\rho({\vec r},t)
 \]
 being the spatial density of probability in the nonrelativistic quantum mechanics or the spatial charge density for   the Dirac particle. We will no longer use the index $0$ at ${\vec r}.$
 

 Thus the change of spatial density  $\rho({\vec r},t)$ with the time 
combined with the Liouville -- von
Neumann -- Wigner equation \eqref{x2} leads to the relation
 \[
\frac{\partial}{\partial t} \sum_{m,n=0}^{s} \int_{{\mathbb R}^3} d\vec{p} \,W(\vec{p},\vec{r},\phi_{m},n, t) + \sum_{m,n=0}^{s}
\int_{{\mathbb R}^3} d\vec{p} \,\{W(\vec{p},\vec{r},\phi_{m},n,t), H(\vec{p},\vec{r},\phi_{m},n,t)\}_M
 =0.
 \]
Comparing this formula with the continuity equation \eqref{01} we see that
 the term
  \\ $
\sum_{m,n=0}^{s}
 \int_{{\mathbb R}^3} d\vec{p} \, \{W(\vec{p},\vec{r},\phi_{m},n,t), H(\vec{p},\vec{r},\phi_{m},n,t)\}_M
 $
 must represent the element $ {\rm div} \vec{j}(\vec{r},t).$ 
 
 \subsection{The nonrelativistic current density of probability }
 \label{subsec51}
 
First let us look at a nonrelativistic 
system, whose Hamilton operator is of the form \eqref{dod1} with $\hat{T}=\frac{\hat{\vec p\,}^2}{2M}$ and the number of possible internal states $s+1=2.$
 In order to construct the corresponding Hamilton function on the quantum
phase space ${\mathbb R}^6 \times \Gamma^2$ we need to apply the correspondence rule \eqref{2321}.

We denote two states referring to the internal degree of freedom  by $|0\rangle $ and $|1\rangle.$ They are chosen as eigenstates of 
the internal potential operator $\hat{V}_{\rm int}.$ Thus in the basis $\{|1\rangle , \, |0\rangle \}$ 
\be
\label{nn1}
\hat{V}_{\rm int}= \left[
\begin{array}{cc}
V_{11} & 0 \\
0 & V_{00}
\end{array}
\right]
\ee
where $V_{11}$ and $V_{00}$ are some real numbers.

Using this set of vectors as a basis
of the Hilbert space ${\mathbb C}^2$ we find that the elements
\[
|\phi_0 \rangle= \frac{1}{\sqrt{2}} \left( |0 \rangle + |1 \rangle \right) \;\;\; , \;\;\; |\phi_1 \rangle= \frac{1}{\sqrt{2}} \left( |0
\rangle - |1 \rangle \right) .
\]
Therefore the operators
\be
\label{n1}
\hat{n}=|1 \rangle \langle 1|\;\;\; , \;\;\; \hat{\phi}= \pi |\phi_1 \rangle \langle \phi_1 |
\ee
are proportional to the projection operators on the directions $|1 \rangle $ and $|\phi_1 \rangle$ respectively. From \eqref{26} we
obtain that the Schwinger operators are
\be
\label{n2}
\widehat{V}= |0 \rangle \langle 0 | - |1\rangle \langle 1| \;\;\; {\rm and } \;\;\; \widehat{R}= 
|0 \rangle \langle 1 | + |1 \rangle \langle 0 |.
\ee
Thus we see that the family of unitary operators $\widehat{D}(k,l)$ (see \eqref{29}) consists of four elements
\bea
\label{n3}
\widehat{D}(0,0)= \hat{\mathbf 1}=|0 \rangle \langle 0 | + |1\rangle \langle 1|, & \widehat{D}(1,0)= \widehat{R}=|0 \rangle \langle 1 |
+ |1 \rangle \langle 0 | \nonumber \\
\widehat{D}(0,1)= \widehat{V}= |0 \rangle \langle 0 | - |1\rangle \langle 1|, & \widehat{D}(1,1)= -i \widehat{R} \widehat{V}= i |0
\rangle \langle 1 | -i |1 \rangle \langle 0 | 
\eea
and the discrete part of the Stratonovich -- Weyl quantiser is determined by
\bea
\label{nn5}
\hat{\Omega}[(-1)^{kl}](\phi_0,0)&=&\frac{1}{2} \Big( 2|0 \rangle \langle 0 | +(1-i)  |0
\rangle \langle 1 | +(1+i) |1 \rangle \langle 0 | \Big),  \nonumber \\
\hat{\Omega}[(-1)^{kl}](\phi_1,0)&=&\frac{1}{2} \Big( 2|0 \rangle \langle 0 | +(-1+i)  |0
\rangle \langle 1 | +(-1-i) |1 \rangle \langle 0 | \Big),  \nonumber \\
\hat{\Omega}[(-1)^{kl}](\phi_0,1)&=&\frac{1}{2} \Big( 2|1 \rangle \langle 1 | +(1+i)  |0
\rangle \langle 1 | +(1-i) |1 \rangle \langle 0 | \Big),  \nonumber \\
\hat{\Omega}[(-1)^{kl}](\phi_1,1)&=&\frac{1}{2} \Big( 2|1 \rangle \langle 1 | +(-1-i)  |0
\rangle \langle 1 | +(-1+i) |1 \rangle \langle 0 | \Big).
\eea
The continuous part of the Stratonovich -- Weyl quantiser equals
\be
\label{nn51}
\hat{\Omega}[1](\vec{p},\vec{r})=\hbar^3  \int\limits_{\mathbb{R}^{3}}\exp\{-i\vec{\lambda}\cdot
\vec{p}\,\}\Big|\vec{r}-\frac{\hbar\vec{\lambda}}{2}\Big> d\vec{\lambda} \,\Big< \vec{r}+\frac{\hbar\vec{\lambda}}{2}\Big|.
\ee
 
From the rule \eqref{2321} applied to the nonrelativistic Hamilton operator \eqref{dod1} with the internal potential \eqref{nn1} we get
\bea
\label{x3}
H({\vec p},\vec{r}, \phi_{0},0 )  =  \frac{{\vec p\,}^2}{2M} + V_{00} \cdot V({\vec{r}}\,), &&
H({\vec p},\vec{r}, \phi_{1},0 )  =  \frac{{\vec p\,}^2}{2M} +V_{00} \cdot V({\vec{r}}\,), \nonumber \\
H({\vec p},\vec{r}, \phi_{0},1 )  =  \frac{{\vec p\,}^2}{2M} +V_{11} \cdot V({\vec{r}}\,), &&
H({\vec p},\vec{r}, \phi_{1},1 )  =  \frac{{\vec p\,}^2}{2M} +V_{11} \cdot V({\vec{r}}\,).
\eea

 Using linearity of the correspondence \eqref{231} we conclude that 
 \[
  \sum_{m,n=0}^1 \int_{{\mathbb R}^3} d\vec{p} \, \{W(\vec{p},\vec{r},\phi_{m},n,t), V_{n\,n} \cdot V({\vec{r}}\,) \}_M=
    0.
 \]  
  Therefore 
 the nonrelativistic current density of probability equals
  \be
  \label{1.5}
  \vec{j}(\vec{r},t) = \frac{1}{M} \sum_{m,n=0}^1 \int_{{\mathbb R}^3} d\vec{p} \,{\vec p\,} \,W(\vec{p},\vec{r},\phi_{m},n,t).
  \ee
  A straightforward consequence of formula \eqref{1.5} is the  fact that the average value of momentum is related to the current density of probability by the integral
$
\langle \vec{p}(t) \rangle = M \int_{{\mathbb R}^3} d \vec{r} \; \vec{j}(\vec{r},t).
$

 \subsection{The relativistic current density  for the $1$ -- D Dirac equation}
 
This subsection contains derivation of the current density for the  $1$ -- D Dirac equation in the phase space quantum mechanics. The first part of construction is devoted to building the phase space counterpart of the  Hamilton operator
\[
\hat{H} = c \hat{p }  (| 0 \rangle \langle 1 | + | 1 \rangle \langle 0 | ) + Mc^2  (| 1 \rangle \langle 1| - | 0 \rangle \langle 0 | )
\]
\be
\label{n5}
+ \hat{V}(x)   \Big(V_{11}| 1 \rangle \langle 1| +V_{10} | 1 \rangle \langle 0|+ V_{01}| 0 \rangle \langle 1 |+V_{00}| 0 \rangle \langle 0 | \Big) .
\ee
 As vectors $|1 \rangle$ and $ |0 \rangle$ we take 
eigenvectors of the Pauli matrix $\sigma_z$  according to formula \eqref{nn4}. Since $\hat{V}_{\rm int}$ is a self -- adjoint operator which in principle does not commute with the operator $\hat{\sigma}_z$, numbers $V_{00}, V_{11} \in {\mathbb R}$ and $V_{01}= \overline{V}_{10}.$

Then we build the Stratonovich -- Weyl quantiser exactly as we did in the
nonrelativistic case in Subsection \ref{subsec51}.
After simple although long calculations based on the rule \eqref{2321} we see that the Hamilton function on the quantum phase space ${\mathbb R} \times \Gamma^2$
equals
\be
\label{nn10}
 \begin{array}{rcl}
H(p,x,\phi_0,0) &=& cp - Mc^2+V(x)\big( V_{00} + \Re((1+i)V_{01}) \big), \\
 H(p,x,\phi_1,0) &=& -cp - Mc^2 +V(x)\big( V_{00} - \Re((1+i)V_{01}) \big), \\
H(p,x, \phi_0,1)& =&  cp + Mc^2 +V(x)\big( V_{11} + \Re((1-i)V_{01}) \big),\\
 H(p,x,\phi_1,1) &=&-cp + Mc^2 +V(x)\big( V_{11} - \Re((1-i)V_{01}) \big),
\end{array} 
\ee
where the symbol $\Re$ denotes the real part.

Analogously like in the nonrelativistic case we observe that only components of the Hamilton function $H(p,x,\phi_m,n)$ containing $p$  contribute to the current density. 
Thus
 the current density for the $1$ -- D Dirac equation equals
\be
\label{1.7}
\vec{j}(x,t)=  q c \int_{\mathbb R} dp \, \left(W(p,x,\phi_0,0,t) + W(p,x,\phi_0,1,t) - W(p,x,\phi_1,0,t)- W(p,x,\phi_1,1,t)\right).
\ee
  
\section{Examples}
\label{sec6}

\setcounter{equation}{0}

In this section we discuss a few illustrative examples of stationary quantum systems characterised by the current density of probability.
The common starting point for them is the star energy eigenvalue equation 
 \be
\label{nn7}
  H(\vec{p},\vec{r},\phi_m,n) * W_E(\vec{p},\vec{r},\phi_m,n)= E W_E(\vec{p},\vec{r},\phi_m,n).
  \ee
  We focus on Hamiltonians which do not depend on time so the respective Wigner eigenfunctions are independent from it. 
For physical reasons we look for real valued functions  $W_E(\vec{p},\vec{r},\phi_m,n)$ assigned to real values of energy $E.$ These requirements imply that for every eigenvalue $E$ the Hamilton function $H(\vec{p},\vec{r},\phi_m,n)$ and the Wigner eigenfunction $W_E(\vec{p},\vec{r},\phi_m,n)$ commute
$
\big\{ H(\vec{p},\vec{r},\phi_m,n), W_E(\vec{p},\vec{r},\phi_m,n)\big\}_{M}=0. 
$

\subsection{A  nonrelativistic free particle with spin $\frac{1}{2}$}

Let us consider a nonrelativistic free particle with the spin $\frac{1}{2}$ living in the $3$ -- D space.  
As the internal states of the particle $|1 \rangle, |0 \rangle$ we choose the eigenstates of the third component of spin applying the convention \eqref{nn4}. The  system is modelled on the grid ${\mathbb R}^6 \times \Gamma^{2}.$

In this case all four components of the Hamilton function are equal to 
\[
H(\vec{p},\vec{r}, \phi_{0},0 )  = H(\vec{p},\vec{r}, \phi_{1},0 )=  H(\vec{p},\vec{r}, \phi_{0},1 )= H(\vec{p},\vec{r}, \phi_{1},1 ) = \frac{\vec{p}^{\;2}}{2M}
\]
and the eigenvalue equation \eqref{nn7} is of the form
\be
\label{nn8}
\left(\frac{\vec{p}^{\;2}}{2M}*W_E \right)(\vec{p},\vec{r},\phi_m,n)= \frac{\vec{p}^{\;2}}{2M}W_E(\vec{p},\vec{r},\phi_m,n) +\frac{i \hbar \vec{p}}{2M} \cdot \nabla W_E(\vec{p},\vec{r},\phi_m,n)
-  \frac{\hbar^2}{8M} \Delta W_E(\vec{p},\vec{r},\phi_m,n)
\ee
\[
= E \,W_E(\vec{p},\vec{r},\phi_m,n), \;\;\; m,n=0,1.
\]
The operators $\nabla$ and $\Delta$ act exclusively on spatial coordinates.

Please notice that we deal with four conditions indexed by the parametres $m$ and $n$ but in this example the eigenvalue formula does not mix terms with different values of   $\phi_m$ and $n.$

 As the eigenstates of energy  in the domain ${\mathbb R}^6$ we choose eigenstates of the momentum function $\vec{p}.$ Assuming that   
 $E= \frac{{\tt p}_x^2+{\tt p}_y^2+{\tt p}_z^2}{2M}$ we obtain 
\be
\label{nn9}
W_{{\tt p}_x\, {\tt p}_y\, {\tt p}_z} (\vec{p},\vec{r},\phi_m,n)= \frac{1}{(2 \pi \hbar)^{3}} C_{mn} \delta(p_x- {\tt p}_x) \delta(p_y- {\tt p}_y) \delta(p_z- {\tt p}_z) \;\;\; m,n=0,1
\ee
and the coefficients $C_{00}, C_{01},C_{10},C_{11}$ are some real numbers fulfilling the conditions  
\[
C_{00}+ C_{01}+C_{10} +C_{11}=1\;, \; C_{00}+ C_{01} \geq 0 \;, \; C_{10}+ C_{11} \geq 0
\;, \; C_{00}+ C_{10} \geq 0 \;, \; C_{01}+ C_{11} \geq 0
\]
arising from the fact that the respective sums represent marginal distributions. They need not be positive.
As expected, the Wigner function \eqref{nn9} does not depend on the position $\vec{r}.$ 

Eliminating the degeneration with respect to the spin we notice that the components of the upper spin Wigner function are
\[
W_{\uparrow \, {\tt p}_x\, {\tt p}_y\, {\tt p}_z} (\vec{p},\vec{r},\phi_0,0)= W_{\uparrow \, {\tt p}_x\, {\tt p}_y\, {\tt p}_z} (\vec{p},\vec{r},\phi_1,0)=0,
\]
\be
\label{y1}
W_{\uparrow \, {\tt p}_x\, {\tt p}_y\, {\tt p}_z} (\vec{p},\vec{r},\phi_0,1)= W_{\uparrow \, {\tt p}_x\, {\tt p}_y\, {\tt p}_z} (\vec{p},\vec{r},\phi_1,1)=  \frac{1}{2(2 \pi \hbar)^{3}}  \delta(p_x- {\tt p}_x) \delta(p_y- {\tt p}_y) \delta(p_z- {\tt p}_z)
\ee
and the down spin Wigner energy eigenfunction is characterised by 
\[
W_{\downarrow \, {\tt p}_x\, {\tt p}_y\, {\tt p}_z} (\vec{p},\vec{r},\phi_0,0)= W_{\downarrow \, {\tt p}_x\, {\tt p}_y\, {\tt p}_z} (\vec{p},\vec{r},\phi_1,0)=  \frac{1}{2(2 \pi \hbar)^{3}}  \delta(p_x- {\tt p}_x) \delta(p_y- {\tt p}_y) \delta(p_z- {\tt p}_z)
\]
\be
\label{y2}
W_{\downarrow \, {\tt p}_x\, {\tt p}_y\, {\tt p}_z} (\vec{p},\vec{r},\phi_0,1)= W_{\downarrow \, {\tt p}_x\, {\tt p}_y\, {\tt p}_z} (\vec{p},\vec{r},\phi_1,1)=0.
\ee
Wigner functions \eqref{y1} and \eqref{y2}, as representing unbound states, are not normalisable.

The current density of probability \eqref{1.5} for the Wigner function 
$W_{\uparrow{\tt p}_x\, {\tt p}_y\, {\tt p}_z} (\vec{p},\vec{r},\phi_m,n)$
as well as $W_{\downarrow{\tt p}_x\, {\tt p}_y\, {\tt p}_z} (\vec{p},\vec{r},\phi_m,n)$
is constant in  space and at every point  equals
\[
\vec{j}_{\uparrow}(\vec{r}\,)=\vec{j}_{\downarrow}(\vec{r}\,)= \frac{1}{(2 \pi \hbar)^3} \frac{\vec{\tt p}}{M}.
\]

\subsection{A $1$ -- D free Dirac particle}
This example presents the eigenstates of the $1$ -- D free Dirac particle.
Since there is  no potential energy, the Hamilton function \eqref{nn10} reduces to 
\be
\label{nn11}
 \begin{array}{rl}
H(p,x,\phi_0,0) = cp - Mc^2, &
H(p,x, \phi_0,1) =  cp + Mc^2
 , \\
 H(p,x,\phi_1,0) = -cp - Mc^2,&
 H(p,x,\phi_1,1) =-cp + Mc^2 .
\end{array} 
\ee
The explicit form of the eigenvalue equation \eqref{nn7} consisting of four different formulas is really long so as an illustration we present  one of them indexed by the values of discrete variables $m=1,\; n=1$:
\[
\frac{1}{2}\Big(  Mc^2 W_E(p,x,\phi_1,1) - cp  W_E(p,x,\phi_1,1)                                                                                                                                           - cp  W_E(p,x,\phi_1,0) +Mc^2 W_E(p,x,\phi_0,1) \Big)
\]
\[
+\frac{i}{2}\Big(  Mc^2 W_E(p,x,\phi_1,0) + cp  W_E(p,x,\phi_0,1)                                                                                                                                           - cp  W_E(p,x,\phi_0,0) -Mc^2 W_E(p,x,\phi_0,0) \Big)
\]
\[
+\frac{i\hbar}{4} \Big(c \frac{\partial W_E(p,x,\phi_1,1)}{\partial x}+ c \frac{\partial W_E(p,x,\phi_1,0)}{\partial x}
\Big)
+\frac{\hbar}{4} \Big(c \frac{\partial W_E(p,x,\phi_0,1)}{\partial x}- c \frac{\partial W_E(p,x,\phi_0,0)}{\partial x}\Big)
\]
\be
\label{nn30}
=E W_E(p,x,\phi_1,1).
\ee
In order to eliminate degeneration we parametrise the solutions by the momentum ${\tt p}$ and the sign of energy. Therefore $E_{\pm} = \pm \sqrt{c^2 {\tt p}^2 + M^2 c^4  }.$

Since the Wigner eigenfunction of the momentum $p$ does not depend on $x,$ the expression \eqref{nn30} does not contain any terms standing at $\hbar.$ Moreover, as the Wigner eigenfunction is real, the formula \eqref{nn30} divides in two linear functional equations.

 The  Wigner  energy eigenfunction of particle (+) as well as antiparticle (-) is proportional to the Dirac delta and has the form 
\[ 
\left\{
\begin{array}{rcl}
W_{{\tt p} \pm}(p,x,\phi_1,1)& =& \frac{1}{4 \pi \hbar} \delta(p- {\tt p}) \frac{c{\tt p}(c{\tt p}-E_{\pm}+ Mc^2)}{(c{\tt p})^2+(E_{\pm}- Mc^2)^2}
,
\vspace{0.2cm}\\
W_{{\tt p} \pm}(p,x,\phi_1,0)&= & \frac{1}{4 \pi \hbar}\delta(p- {\tt p}) \frac{(E_{\pm}- Mc^2)(E_{\pm}- Mc^2-c{\tt p})}{(c{\tt p})^2+(E_{\pm}- Mc^2)^2},
\vspace{0.2cm}\\
W_{{\tt p} \pm}(p,x, \phi_0,1)&=&\frac{1}{4 \pi \hbar}\delta(p- {\tt p}) \frac{c{\tt p}(c{\tt p}+E_{\pm}- Mc^2)}{(c{\tt p})^2+(E_{\pm}- Mc^2)^2},
\vspace{0.2cm}\\
W_{{\tt p} \pm}(p,x,\phi_0,0)& =& \frac{1}{4 \pi \hbar} \delta(p- {\tt p}) \frac{(E_{\pm}- Mc^2)(c{\tt p}+E_{\pm}- Mc^2)}{(c{\tt p})^2+(E_{\pm}- Mc^2)^2}.
\end{array} \right.
\]
One of components of Wigner function for the particle and one for the antiparticle is negative.
 
Substituting this result into definition \eqref{1.7} we obtain that the relativistic  current density is stationary and equals
\be
\label{nn12}
\vec{j}_{{\tt p} \pm}(x)= \pm  \frac{q c^2 \vec{\tt p}}{2 \pi \hbar \sqrt{c^2 {\tt p}^2+M^2 c^4}}
\ee
as expected.

In the nonrelativistic limit $Mc^2 \gg |c {\tt p}|$ the Wigner function of particle reduces to 
\[
W_{{\tt p} +}(p,x,\phi_1,1)=W_{{\tt p} +}(p,x,\phi_0,1)= \frac{1}{4 \pi \hbar}\delta(p- {\tt p})\;\; , \;\; W_{{\tt p} +}(p,x,\phi_1,0)= W_{{\tt p} +}(p,x,\phi_0,0)=0
\]
and the Wigner function of antiparticle equals
\[
W_{{\tt p} -}(p,x,\phi_1,1)=W_{{\tt p} -}(p,x,\phi_0,1)=0\;\; , \;\; W_{{\tt p} -}(p,x,\phi_1,0)= W_{{\tt p} -}(p,x,\phi_0,0)= \frac{1}{4 \pi \hbar}\delta(p- {\tt p}).
\]
These formulas are analogous to the ones derived for the free nonrelativistic particle with the spin $\frac{1}{2}.$
\subsection{A nonrelativistic scattering of spin $\frac{1}{2}$ particles on  the step potential }
\label{subsection6.3}

In this paragraph we discuss the $3$ -- D  nonrelativistic motion  of quantum particles with spin $\frac{1}{2}$ with the step barrier  $V= V(x) \otimes \hat{\mathbf 1},$
where
\begin{equation}
\label{nn23.0}
    V(x) = \begin{cases} 
     0 & \mbox{for } x < 0,
     \\ V_0 & \mbox{for } x \geq 0 
    \end{cases}
\end{equation}
and $V_0>0.$ Because of the shape of the barrier we assume that particles are moving exclusively in the $x$ direction. Thus all considerations are done like for $1$ -- D case but  the spin exists. This kind of potential does not interact with the internal angular momentum. We focus on the situation when the energy of particles in the incoming beam is greater than $V_0.$ 

In principle the problem can be solved exclusively in frames of phase space quantum mechanics. However, since the final result is well known and in order to solve the energy eigenvalue equation  \eqref{nn7} we would have to apply the integral definition \eqref{610} of the star product, we prefer to find the Wigner energy eigenfunction directly from  the formula \eqref{2322}. 

The beam of scattered particles consists of molecules of the fixed momentum ${\tt p}= \sqrt{2ME}, \; E>V_0$ before the barrier ($x<0$) and ${ \tilde{\tt p}}=\sqrt{2M(E-V_0)}$ over the barrier ($x>0$).
Therefore the respective wave function is given by the expression
\[
\Psi_E(x)= \left[ 
\begin{array}{c}A_1 \\A_0
\end{array}
\right] \psi_E(x) =  \frac{Y(-x)}{2}
\left[ 
\begin{array}{c}A_1 \\A_0
\end{array}
\right] 
\left( \left( 1+ \frac{\tilde{\tt p}}{\tt p}\right) \exp\left( \frac{i{\tt p}x}{\hbar}\right)
+ \left( 1- \frac{\tilde{\tt p}}{\tt p}\right) \exp\left(- \frac{i {\tt p}x}{\hbar}\right) \right)
\]
\be
\label{nn12}
+Y(x) 
\left[
\begin{array}{c}A_1 \\ A_0
\end{array}
\right]  \exp\left( \frac{i\tilde{\tt p}x}{\hbar}\right), \;\;\; |A_1|^2 + |A_0|^2=1.
\ee
Thus from the correspondence rule \eqref{2322} we get that the components of the Wigner function are
\be
\label{nn13} 
\left\{
\begin{array}{rcl}
W_E(p,x,\phi_0,0)& =& \frac{1}{2 \sqrt{2 \pi} \hbar} (|A_0|^2 + \Re((1+i)\overline{A_0}A_1) ){\cal F}_{\xi} \left[ \overline{\psi_E}\left( x+
\frac{{\xi}}{2}\right) \psi_E\left( x- \frac{{\xi}}{2}\right) \right] \left( \frac{p}{\hbar}\right),
\vspace{0.2cm}\\
W_E(p,x,\phi_1,0)&= & \frac{1}{2 \sqrt{2 \pi} \hbar} (|A_0|^2 - \Re((1+i)\overline{A_0}A_1)){\cal F}_{\xi} \left[ \overline{\psi_E}\left( x+
\frac{{\xi}}{2}\right) \psi_E\left( x- \frac{{\xi}}{2}\right) \right] \left( \frac{p}{\hbar}\right),
\vspace{0.2cm}\\
W_E(p,x, \phi_0,1)&=&\frac{1}{2 \sqrt{2 \pi} \hbar} (|A_1|^2 + \Re((1-i)\overline{A_0}A_1)){\cal F}_{\xi} \left[ \overline{\psi_E}\left( x+
\frac{{\xi}}{2}\right) \psi_E\left( x- \frac{{\xi}}{2}\right) \right] \left( \frac{p}{\hbar}\right),
\vspace{0.2cm}\\
W_E(p,x,\phi_1,1)& =& \frac{1}{2 \sqrt{2 \pi} \hbar} (|A_1|^2 - \Re((1-i)\overline{A_0}A_1)){\cal F}_{\xi} \left[ \overline{\psi_E}\left( x+
\frac{{\xi}}{2}\right) \psi_E\left( x- \frac{{\xi}}{2}\right) \right] \left( \frac{p}{\hbar}\right)
\end{array} 
\right.
\ee
where the Fourier transform is defined as
 \[
  {\cal F}[\phi(z)](t)={\cal F}_z[\phi](t)   := \frac{1}{\sqrt{2 \pi}} \int_{{\mathbb R}} \phi(z) \exp (i z t) dz.
  \]

The fact that the relationship between the wave function $\psi_E(x)$ and the Wigner function \eqref{nn13} is based on the Fourier transform of the product of translated functions, leads to an interesting observation that the value of the Wigner function at a fixed point $(p_0,x_0)$ depends on the values of  wave function $\psi_E(x)$ in the whole space. This effect is caused by different ways in which the wave function and the Wigner function encode the information about the momentum.

Let us  focus at the component 
\[
 \frac{1}{ \sqrt{2 \pi} \hbar} {\cal F}_{\xi} \left[ \overline{\psi_E}\left( x+
\frac{{\xi}}{2}\right) \psi_E\left( x- \frac{{\xi}}{2}\right) \right] \left( \frac{p}{\hbar}\right)=
\]
\[
 \left( 1+\frac{\tilde{\tt p}}{\tt p}\right) \cos \left( \frac{({\tt p}- \tilde{\tt  p})x}{\hbar}\right) \delta(2 p-{\tt p}- \tilde{\tt p})
+ \left( 1-\frac{\tilde{\tt p}}{\tt p}\right) \cos \left( \frac{({\tt p}+ \tilde{\tt p})x}{\hbar}\right) \delta(2 p+{\tt p}- \tilde{\tt p})
\]
\[
+ \frac{1}{ \pi} \left( 1-\frac{\tilde{\tt p}}{\tt p}\right) \sin \left( \frac{({\tt p} + \tilde{\tt p})x-(2p+{\tt p}-\tilde{\tt p})|x|}{\hbar}\right) {\rm vp} \frac{1}{2p+{\tt p}-\tilde{\tt p}}
\]
\[
+ \frac{1}{ \pi} \left( 1+\frac{\tilde{\tt p}}{\tt p}\right) \sin \left( \frac{({\tt p} - \tilde{\tt p})x-(2p-{\tt p}-\tilde{\tt p})|x|}{\hbar}\right) {\rm vp} \frac{1}{2p-{\tt p}-\tilde{\tt p}}
\]
\[
-\frac{1}{4 \pi} \left( 1-\frac{\tilde{\tt p}}{\tt p}\right)^2 \sin\left( \frac{2x(p+{\tt p})}{\hbar}\right) \frac{Y(-x)}{p+{\tt p}} 
-\frac{1}{4 \pi} \left( 1+\frac{\tilde{\tt p}}{\tt p}\right)^2 \sin\left( \frac{2x(p-{\tt p})}{\hbar}\right) \frac{Y(-x)}{p-{\tt p}} 
\]
\be
\label{nn14}
-\frac{1}{2 \pi} \left( 1-\left(\frac{\tilde{\tt p}}{\tt p}\right)^2\right) \cos\left( \frac{2{\tt p}x}{\hbar}\right)\sin\left( \frac{2px}{\hbar}\right) \frac{Y(-x)}{p} 
+ \frac{1}{ \pi} \sin\left( \frac{2x(p-\tilde{\tt p})}{\hbar}\right)  \frac{Y(x)}{p-\tilde{\tt p}}
\ee
of the Wigner function \eqref{nn13}. 

The purely transmitted term (the counterpart of the component $Y(x) \exp\left( \frac{i\tilde{\tt p}x}{\hbar}\right)$ of the incident wave function) is represented by the function 
\[
W_{E\,trans}(p,x)= \frac{1}{ \pi} \sin\left( \frac{2x(p-\tilde{\tt p})}{\hbar}\right)  \frac{Y(x)}{p-\tilde{\tt p}},
\]
the incident part equals 
\[
W_{E\,inc}(p,x)=
-\frac{1}{4 \pi} \left( 1+\frac{\tilde{\tt p}}{\tt p}\right)^2 \sin\left( \frac{2x(p-{\tt p})}{\hbar}\right) \frac{Y(-x)}{p-{\tt p}} 
\]
and the reflected one is
\[
W_{E\,ref}(p,x)=
-\frac{1}{4 \pi} \left( 1-\frac{\tilde{\tt p}}{\tt p}\right)^2 \sin\left( \frac{2x(p+{\tt p})}{\hbar}\right) \frac{Y(-x)}{p+{\tt p}}. 
\]
Other parts of expression \eqref{nn14} origin from interference between the incident element of the wave function \eqref{nn12}, its reflected component and the transmitted one under the Fourier transform.

As one can notice easily, the direct application of formula \eqref{1.5} to calculating components of the current density of probability leads to divergent expressions. To avoid this obstacle we introduce a family of auxiliary functions
for which the integral $\int_{\mathbb R} dp$ is convergent (see  \cite{schwartz65}). For example for the transmitted beam we put 
\[
W_{E \, trans}(p,x, \alpha):= \exp \left( - \alpha |p|\right)W_{E\, trans}(p,x), \;\; \alpha >0.
\]
Of course now the integral  $\int_{\mathbb R}pW_{E \, trans}(p,x, \alpha)  dp$ is convergent for every admissible value of the parametre $\alpha.$ We postulate that
\[
\int_{\mathbb R}pW_{E \, trans}(p,x)  dp = \lim_{\alpha \rightarrow 0^+} \int_{\mathbb R} pW_{E \, trans}(p,x, \alpha)dp.
\]
From \eqref{1.5} for the Wigner function \eqref{nn13}  on the grid we get that in the $X$ axis  direction the components of the the current density are
\[
j_{trans}(x)=  \frac{ \tilde{\tt p}}{M}Y(x)\;,\; j_{inc}(x)= \frac{1}{4 } \left( 1+\frac{\tilde{\tt p}}{\tt p}\right)^2  \frac{\tt p}{M}Y(-x),
\]
\be
\label{nn20}
 j_{ref}(x)= -\frac{1}{4 } \left( 1-\frac{\tilde{\tt p}}{\tt p}\right)^2  \frac{\tt p}{M}Y(-x)
\ee
as expected. 

A very interesting question is a contribution of interference terms to spatial density of probability and the current density of probability (see also \cite{tosiek16}). Let us consider  two parts of a   wave function $\psi(x)$
\be
\label{nn21}
\psi_1(x)=Y(x-a_1)Y(b_1-x) \psi(x) \;{\rm and }\; \psi_2(x)= Y(x-a_2)Y(b_2-x) \psi(x),
\ee
where the numbers $a_1, b_1, a_2, b_2$ fulfill the requirements 
\[
  - \infty \leq a_1 < b_1 < a_2 <b_2 \leq \infty.
\]
The contribution of mixture of these two terms \eqref{nn21} to the total Wigner function equals 
\[
W_{12}(p,x)=\frac{2}{ \sqrt{2 \pi} \hbar} \Re \left\{{\cal F}_{\xi} \left[ \overline{\psi}_1\left( x+
\frac{{\xi}}{2}\right) \psi_2\left( x- \frac{{\xi}}{2}\right) \right] \left( \frac{p}{\hbar}\right) \right\}.
\]
But 
\be
\label{nn22}
\forall \;
n \in {\mathcal N} \;\;\forall \; x \in {\mathbb R}
\int_{\mathbb R} dp \,p^n \, W_{12}(p,x)=0 \;\;\; 
\ee
so the interference terms do not influence neither the spatial density of probability nor the current density. Moreover, this observation does not depend on the shape of the potential barrier. 
And for any arbitrary analytical function $f(p,x)$ on ${\mathbb R}^2$ the term $W_{12}(p,x)$ does not contribute to the mean value $\langle f(p,x) \rangle.$

The role of the interference component of the Wigner function originated from an incident beam and the reflected one requires a separate analysis because the incident wave and the reflected one are defined on the same domain. As before, since an internal degree  is not involved in calculations, we assume that
\be
\label{nn22}
\psi_{E\, inc}(x) \sim Y(b-x) \exp \left( \frac{i {\tt p}x}{\hbar}\right)\; , \; 
\psi_{E\, ref}(x) \sim Y(b-x) \exp \left(- \frac{i {\tt p}x}{\hbar}\right), \;\; b \in {\mathbb R}.
\ee
Therefore for arbitrary $x<b$ one can see that
\[
\forall  \;
n \in {\mathcal N} / \{ 0 \} \;\; 
\int_{\mathbb R} dp  \,p^n \, W_{E \,inc \, ref}(p,x)=0
\]
so this part of the Wigner function does not influence  the current density of probability.

\subsection{The $1$ -- D Klein paradox}

This last example presents  a discussion of motion of a $1$ -- D Dirac particle in the step potential \eqref{nn23.0}. 
From \eqref{n5} we can see that its Hamilton operator equals
\be
\label{nn24} 
\hat{H} = c \hat{p }  (| 0 \rangle \langle 1 | + | 1 \rangle \langle 0 | ) + Mc^2  (| 1 \rangle \langle 1| - | 0 \rangle \langle 0 | )
+ \hat{V}(x)   \Big(| 1 \rangle \langle 1| +| 0 \rangle \langle 0 | \Big) 
\ee
with the potential $V(x)$ given by formula \eqref{nn23.0}.

An amazing effect can be observed  for particles when  $Mc^2 < E $ and $E+ Mc^2 <V_0.$ Instead of the probability tending rapidly to $0$ inside the barrier we obtain an oscillating solution representing a wave running to infinity. The solution of $1$ -- D Dirac equation is the function 
\[
\Psi_{E}(x)= Y(-x)    \left[ \begin{array}{c}
    \frac{c{\tt p}}{E- Mc^2} \\ 1
    \end{array} \right]  \exp\left( \frac{i{\tt p}x}{ \hbar } \right) +  Y(-x)  N_{ref}    \left[ \begin{array}{c}
    \frac{-c{\tt p}}{E- Mc^2} \\ 1
    \end{array} \right]  
    \exp\left(- \frac{i{\tt p}x}{ \hbar } \right) 
\]
\be
\label{nn29}
 + Y(x)  N_{trans}  \left[  \begin{array}{c}
    \frac{c \tilde{\tt p}}{E-V_0 - Mc^2} \\ 1
    \end{array} \right]
    \exp\left( \frac{i \tilde{\tt p}x}{ \hbar } \right)
    \ee
with
\[
E = \sqrt{{\tt p}^2 c^2 + M^2 c^4}= - \sqrt{\tilde{\tt p}^2 c^2 + M^2 c^4}+ V_0.
\]
The coefficients ${\tt p}, \; \tilde{\tt p} $ are positive. They fulfill the inequality 
 $\tilde{\tt p}> {\tt p} $ if $V_0>2E.$ 
 
The continuity of function $\Psi_E(x)$ at $x=0$ implies that the coefficient is equal to
\setcounter{orange}{1} \renewcommand{\theequation}{\arabic{section}.\arabic{equation}\theorange} 
\be
\label{nn23.1}
N_{trans}= 1 + \frac{E}{Mc^2}+ \frac{Mc^2}{V_0} + \frac{\sqrt{E^2 - M^2 c^4} \sqrt{(E-V_0)^2-M^2c^4}}{Mc^2 V_0} - \frac{E^2}{Mc^2 V_0}
\ee
and
\addtocounter{orange}{1} \addtocounter{equation}{-1} 
\be
\label{nn23.2}
N_{ref}=N_{trans}-1.
\ee
 \renewcommand{\theequation}{\arabic{section}.\arabic{equation}} 
  The value of the  coefficient $N_{trans}$ changes from $2$ for $V_0= E + Mc^2$ till $1 +\frac{E}{Mc^2}+ \frac{\sqrt{E^2-M^2 c^4}}{Mc^2}$ for $V_0 \rightarrow \infty.$

 In  Subsection \ref{subsection6.3} we proved, that only elements $W_{inc}(p,x), \, W_{ref}(p,x), \, W_{trans}(p,x)$ of the Wigner function contribute in the current density. 
Thus exclusively the elements
\setcounter{orange}{1} \renewcommand{\theequation}{\arabic{section}.\arabic{equation}\theorange} 
\be
\label{nn24} 
\left\{
\begin{array}{rcl}
W_{inc}(p,x,\phi_0,0)& =& -\frac{1}{2 \pi} \frac{E-Mc^2+c{\tt p}}{E-Mc^2}  \sin\left( \frac{2x(p-{\tt p})}{\hbar}\right) \frac{Y(-x)}{p-{\tt p}} ,
\vspace{0.2cm}\\
W_{inc}(p,x,\phi_1,0)&= & -\frac{1}{2 \pi} \frac{E-Mc^2-c{\tt p}}{E-Mc^2}  \sin\left( \frac{2x(p-{\tt p})}{\hbar}\right) \frac{Y(-x)}{p-{\tt p}} ,
\vspace{0.2cm}\\
W_{inc}(p,x, \phi_0,1)&=&-\frac{1}{2 \pi} \frac{c {\tt p}(E-Mc^2+c{\tt p})}{(E-Mc^2)^2}  \sin\left( \frac{2x(p-{\tt p})}{\hbar}\right) \frac{Y(-x)}{p-{\tt p}} ,
\vspace{0.2cm}\\
W_{inc}(p,x,\phi_1,1)& =& -\frac{1}{2 \pi}\frac{-c {\tt p}(E-Mc^2-c{\tt p})}{(E-Mc^2)^2}  \sin\left( \frac{2x(p-{\tt p})}{\hbar}\right) \frac{Y(-x)}{p-{\tt p}} 
\end{array} 
\right.,
\ee
\addtocounter{orange}{1} \addtocounter{equation}{-1} 
\be
\label{nn25} 
\left\{
\begin{array}{rcl}
W_{ref}(p,x,\phi_0,0)& =& -\frac{(1-N_{trans})^2}{2 \pi} \frac{E-Mc^2-c{\tt p}}{E-Mc^2}  \sin\left( \frac{2x(p+{\tt p})}{\hbar}\right) \frac{Y(-x)}{p+{\tt p}} ,
\vspace{0.2cm}\\
W_{ref}(p,x,\phi_1,0)&= & -\frac{(1-N_{trans})^2}{2 \pi} \frac{E-Mc^2+c{\tt p}}{E-Mc^2}  \sin\left( \frac{2x(p+{\tt p})}{\hbar}\right) \frac{Y(-x)}{p+{\tt p}} ,
\vspace{0.2cm}\\
W_{ref}(p,x, \phi_0,1)&=&-\frac{(1-N_{trans})^2}{2 \pi} \frac{-c {\tt p}(E-Mc^2-c{\tt p})}{(E-Mc^2)^2}  \sin\left( \frac{2x(p+{\tt p})}{\hbar}\right) \frac{Y(-x)}{p+{\tt p}} ,
\vspace{0.2cm}\\
W_{ref}(p,x,\phi_1,1)& =& -\frac{(1-N_{trans})^2}{2 \pi}\frac{c {\tt p}(E-Mc^2+c{\tt p})}{(E-Mc^2)^2}  \sin\left( \frac{2x(p+{\tt p})}{\hbar}\right) \frac{Y(-x)}{p+{\tt p}} 
\end{array} 
\right.
\ee
\addtocounter{orange}{1} \addtocounter{equation}{-1} 
and
\be
\label{nn26} 
\left\{
\begin{array}{rcl}
W_{trans}(p,x,\phi_0,0)& =& \frac{N_{trans}^2}{2 \pi} \frac{E-V_0-Mc^2+c\tilde{\tt p}}{E-V_0-Mc^2}  \sin\left( \frac{2x(p-\tilde{\tt p})}{\hbar}\right) \frac{Y(x)}{p-\tilde{\tt p}} ,
\vspace{0.2cm}\\
W_{trans}(p,x,\phi_1,0)&= & \frac{N_{trans}^2}{2 \pi} \frac{E-V_0-Mc^2-c \tilde{\tt p}}{E-V_0-Mc^2} \sin\left( \frac{2x(p-\tilde{\tt p})}{\hbar}\right) \frac{Y(x)}{p-\tilde{\tt p}} ,
\vspace{0.2cm}\\
W_{trans}(p,x, \phi_0,1)&=&\frac{N_{trans}^2}{2 \pi} \frac{c \tilde{\tt p}(E-V_0-Mc^2+c\tilde{\tt p})}{(E-V_0-Mc^2)^2}  \sin\left( \frac{2x(p-\tilde{\tt p})}{\hbar}\right) \frac{Y(x)}{p-\tilde{\tt p}} ,
\vspace{0.2cm}\\
W_{trans}(p,x,\phi_1,1)& =& \frac{N_{trans}^2}{2 \pi} \frac{-c \tilde{\tt p}(E-V_0-Mc^2-c\tilde{\tt p})}{(E-V_0-Mc^2)^2}  \sin\left( \frac{2x(p-\tilde{\tt p})}{\hbar}\right) \frac{Y(x)}{p-\tilde{\tt p}} 
\end{array} 
\right.
\ee
\renewcommand{\theequation}{\arabic{section}.\arabic{equation}} 
calculated with the use of correspondence rule \eqref{2322}
give contribution to the current density.

From the formula \eqref{1.7} we obtain that the charge current densities 
\[
j_{inc}(x)= Y(-x)\frac{2c^2q {\tt p}}{E-Mc^2}\;\; , \;\;j_{ref}(x)= -Y(-x)\frac{2c^2q {\tt p}(1-N_{trans})^2}{E-Mc^2},
\]
\be
\label{nn27}
j_{trans}(x)= Y(x)\frac{2c^2q \tilde{\tt p}N_{trans}^2}{E-V_0-Mc^2}.
\ee
Please notice that the transmitted current density is negative.
Thus the transmission coefficient $T$
 may exceed 1. 
 
 \begin{figure}[H]\centering
\label{rys2}
\includegraphics[width=120mm]{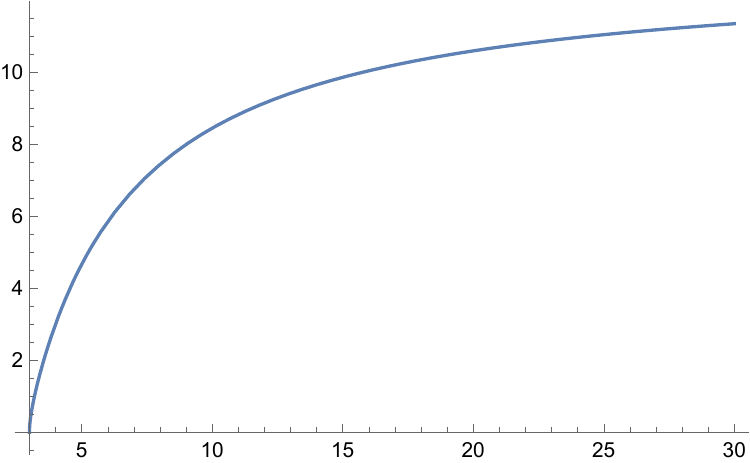} 
\caption{The transmission coefficient as a function of $V_0$ for $V_0>E+Mc^2.$  \\
\hspace*{1.75cm} At the picture $ E=2,  \; M=1, \; c=1$.}
\end{figure} 
 
 The fact that the transmitted current is negative, changes the standard relation between the reflection coefficient and the transmission coefficient.
Now 
\[
R-T=1.
\]
This result is called  {\it the Klein paradox}. Its   detailed discussion  can be found in \cite{Kl29, AH81, WG90, PS98}.



\end{document}